\RequirePackage{fix-cm}
\documentclass[smallextended]{svjour3}       
\smartqed  
\usepackage{graphicx}
\usepackage{listings}
\usepackage{url}
\usepackage{amsmath}
\usepackage{booktabs}
\usepackage{array}
\usepackage{xcolor}
\usepackage{caption}
\usepackage{adjustbox}
\usepackage{hyperref}
\usepackage{placeins}
\AtBeginDocument{\hypersetup{
    pdftitle={Design Pattern Summariser (DPS): Empirical Software Engineering Study},
    pdfauthor={Najam Nazar; Sameer Sikka; Christoph Treude},
    pdfsubject={Empirical evaluation of design pattern summarisation},
    pdfkeywords={design patterns, summarisation, software engineering, BLEU, ROUGE, NIST, FrugalScore}
}}
\usepackage{multirow}

\lstdefinestyle{JavaStyle}{
    language=Java,
    basicstyle=\ttfamily\footnotesize,
    keywordstyle=\color{blue},
    commentstyle=\color{gray},
    stringstyle=\color{red},
    showstringspaces=false,
    tabsize=4,
    breaklines=true
}
\journalname{Empirical Software Engineering}
\begin{document}
\title{DPS: Design Pattern Summarisation Using Code Features}{
}


\author{Najam Nazar \and Sameer Sikka \and Christoph Treude}


\institute{
    Najam Nazar \at
    School of Computer and Mathematical Sciences, Adelaide University, Australia \\
    Department of Software Systems \& Cybersecurity, Faculty of Information Technology, Monash University, Australia\\
    \email{najam.nazar@adelaide.edu.au, najam.nazar@monash.edu}\\
    ORCID: \href{https://orcid.org/0000-0003-2317-2297}{0000-0003-2317-2297}
    \and
    Sameer Sikka \at
    Faculty of Engineering and Information Technology, The University of Melbourne, Australia \\
    \email{ssikka@student.unimelb.edu.au}\\
    ORCID: \href{https://orcid.org/0009-0000-7796-0008}{0009-0000-7796-0008}
    \and
    Christoph Treude \at
    School of Computing and Information Systems,
    Singapore Management University, Singapore\\
    \email{ctreude@smu.edu.sg}\\
    ORCID: \href{https://orcid.org/0000-0002-6919-2149}{0000-0002-6919-2149}
}

\date{Received: date / Accepted: date}

\maketitle

\begin{abstract}
Automatic summarisation has been used efficiently in recent years to condense texts, conversations, audio, code, and various other artefacts. A range of methods, from simple template-based summaries to complex machine learning techniques---and more recently, large language models---have been employed to generate these summaries. Summarising software design patterns is important because it helps developers quickly understand and reuse complex design concepts, thereby improving software maintainability and development efficiency. However, the generation of summaries for software design patterns has not yet been explored. 

Our approach utilises code features and JavaParser to parse the code and create a JSON representation. Using an NLG library on this JSON representation, we convert it into natural language text that acts as a summary of the code, capturing the contextual information of the design pattern. Our empirical results indicate that the summaries generated by our approach capture the context in which patterns are applied in the codebase. Statistical evaluations demonstrate that our summaries closely align with human-written summaries, as evident from high values in the ROUGE-L, BLEU-4, NIST, and FrugalScore metrics. A follow-up survey further shows that DPS summaries were rated as capturing context better than human-generated summaries. Additionally, a time based task activity shows that summaries increase the time of understanding of design pattern for developer better than when the summaries are not present.

\textcolor{blue}{To our knowledge, this work presents the first automated approach to summarising software design patterns by linking pattern-detection features with natural-language generation. Unlike existing design-pattern detection or code summarisation methods that address these tasks separately, DPS introduces a feature-driven integration that converts detected pattern structures into contextual summaries describing the roles, relationships, and usage intent of each pattern instance.}

\end{abstract}

\section{Introduction}
\label{sec:intro}

Software design patterns are reusable solutions to common software design problems~\cite{kuchana:2004,Gamma:1995}. They provide a way for developers to solve problems in a consistent and efficient manner and to communicate their solutions to other developers~\cite{Nazar:2022}. Design patterns are abstract and conceptual and do not specify how to implement them in a particular programming language or platform~\cite{Linkedin:2024}. Identifying software design patterns is a well-established research area in computer science and programming, and many studies have been conducted, including~\cite{Nazar:2022}, \cite{Zanoni:2015}, and \cite{Thaller:2019}.

Using design patterns in code introduces several challenges. The first challenge is to implement them without a proper understanding of the context or flexibility, which can lead to over-engineering, increased complexity, and reduced code adaptability~\cite{Profocus:2025}. The second challenge is to ensure a correct and consistent implementation, as factors such as software aging, lack of developer experience, and the mutation of patterns into anti-patterns can pose significant issues~\cite{Baumgartner:2019}, \cite{Kermansaravi:2021}. A third challenge is effectively communicating design patterns to other developers. Although patterns are intended to facilitate communication and collaboration by providing a common vocabulary and understanding of software design, the lack of standardised documentation and varying developer interpretations can result in inconsistent implementations, further complicating effective communication~\cite{Wedyan:2020}. Therefore, it is crucial to communicate patterns clearly and explicitly using appropriate diagrams, code comments, documentation, and explaining the rationale and benefits of their use.

To address these challenges, we propose the automatic generation of summaries for design patterns, with a focus on the context in which these patterns are used within the source code. The generation of design pattern summaries offers several advantages to software developers. These include enhanced comprehension and communication of design patterns, improved code maintainability, and the facilitation of effective pattern reuse with minimal customisation in their usage contexts. Additionally, generating summaries of design patterns provides concise explanations and clarifies the relationships between patterns~\cite{Gamma:1995}, thus facilitating communication and decision making between developers working on the same project~\cite{Linkedin:2024}.

Building on existing study on design pattern detection~\cite{Nazar:2022}, where they identified software design patterns using source code features, we leverage these features to generate concise descriptions of common design patterns. We refer to this new approach as the \textbf{Design Pattern Summariser (DPS)}. To achieve this, we first collected a corpus of Java files from open-source GitHub projects that specifically implement design patterns. In total, we collected ten such projects, which we refer to as the \textbf{DPS-Corpus}. We parsed the DPS-Corpus by constructing an abstract syntax tree (AST) using source code features, ultimately generating a JSON representation. Next, we used an updated pattern detection approach based on the existing work~\cite{Nazar:2022} to identify design patterns within the DPS-Corpus and subsequently added or updated the respective pattern information in the JSON representation.

Using the SimpleNLG library\footnote{\url{https://github.com/simplenlg/simplenlg} accessed and verified on 31-03-2025}, we converted the JSON information into natural language summary sentences that capture the context in which design patterns are used. To evaluate the efficacy of our approach, we asked human participants to write summaries of the DPS-Corpus files and compared these with the machine-generated summaries produced by our approach. Human summaries were collected through an online survey involving university staff and post-graduate students with expertise in software engineering. We calculated \textit{BLEU-4}, \textit{NIST}, \textit{ROUGE-L}, and \textit{FrugalScore} statistical metrics to compare the summaries. Additionally, we conducted an online survey to assess the quality of the summaries by asking five quality-related questions. The results indicate that the DPS summaries capture the context of design patterns more effectively than the human summaries.

Our work focuses on summarising nine commonly used software design patterns---comprising three patterns from each category of design patterns: creational, structural, and behavioural. The selected patterns are \textit{Abstract Factory}, \textit{Adapter}, \textit{Decorator}, \textit{Fa\c{c}ade}, \textit{Factory Method}, \textit{Memento}, \textit{Observer}, \textit{Singleton}, and \textit{Visitor}.


\textcolor{blue}{While previous research has either focused on detecting design patterns from source code or on generating summaries of general code elements, our work is the first to bridge these two directions. The novelty of DPS lies in \textbf{(1)} the construction of a feature-enhanced representation that encodes both class-level and method-level context for design-pattern instances, and \textbf{(2)} the generation of pattern-aware natural-language summaries that describe not only structural properties but also the contextual roles of participating classes. This integration enables automated documentation of design patterns, a problem that has not been previously studied.}

The main contributions of our work are as follows:

\begin{itemize}
    \item \textcolor{blue}{We introduce the first approach that integrates feature-based design-pattern detection with automatic natural-language generation to produce pattern-aware summaries describing the roles and relationships within design-pattern instances.}
    \item \textcolor{blue}{We construct a large corpus of 15 educational and industrial open-source Java systems (42,000+ files) containing implementations of nine GoF design patterns.}
    \item \textcolor{blue}{We extend existing feature-based detection techniques by expanding the feature set from 15 to 20 and using these features as structured inputs to a pattern-aware summarisation pipeline.}
    \item \textcolor{blue}{We empirically evaluate the approach using automatic metrics, expert judgement, and a comprehension study, showing that DPS summaries capture design-pattern context comparably or better than human-written summaries and improve developers’ comprehension efficiency.}
\end{itemize}

The paper is organised into the following sections. Section~\ref{sec:intro} introduces the problem, and Section~\ref{sec:context} provides the motivation behind the paper, particularly by discussing what context means. Section~\ref{sec:exper} discusses the experimental setup, focusing on corpus selection and generation, the code features used, the statistical measures for evaluation, and the research questions. Section~\ref{sec:dps} describes our three-phase summariser approach---DPS. Section~\ref{sec:empir} evaluates our work based on statistical measures and survey-based feedback. Section~\ref{sec:discuss} discusses the results as well as the task based activity we performed. Section~\ref{sec:threats} discusses threats to internal and external validity, while Section~\ref{sec:work} provides a description of related work in design pattern detection and code summarisation. Section~\ref{sec:conc} concludes our paper and discusses future directions for this research.

\section{Background:Context}\label{sec:context}

\textcolor{blue}{The notion of context in source code has been interpreted in various ways across the literature~\cite{Mcburney:2014,Makharev:2025}. In this work, we define context as the structural and behavioural information surrounding a code element (e.g., class or method) that is necessary to understand its role and functionality within a design-pattern instance. Specifically, we distinguish two types of context:}

\begin{itemize}
\item \textcolor{blue}{Method-level context, which captures the behaviour of a method through its dependencies, including the methods it invokes and the methods that invoke it.}
\item \textcolor{blue}{Class-level context, which refers to the role a class plays within the pattern (e.g., acting as a Subscriber, Publisher, ConcreteFactory, or Product) and its structural relationships such as inheritance, composition, or interface implementation.}
\end{itemize}

\textcolor{blue}{McBurney et al.~\cite{Mcburney:2014} primarily focused on method-level context, defining it as the purpose of a method within the code in which it exists, including its dependencies and dependents. While this is essential for local comprehension tasks, in the case of design patterns the class-level context becomes even more critical, as it conveys the structural, behavioural, and creational intent of the pattern. For example, knowing whether a class functions as a ConcreteFactory or a Product in the Abstract Factory pattern strongly shapes how developers interpret its purpose and interactions.}

\textcolor{blue}{Our DPS summarisation approach therefore captures both class-level and method-level context, representing them hierarchically: class-level roles and relationships form the high-level pattern narrative, while method-level behaviour is incorporated when it clarifies how these roles collaborate within the pattern.}

The following example, taken from the RefactoringGuru GitHub repository,\footnote{\url{https://github.com/RefactoringGuru/design-patterns-java/blob/main/src/refactoring_guru/abstract_factory/example/factories/GUIFactory.java} accessed and verified on 31-03-2025} illustrates the usage of context with respect to design patterns.

\begin{lstlisting}[style=JavaStyle]
public interface GUIFactory { 
    Button createButton(); 
    Checkbox createCheckbox();
}
\end{lstlisting}

The interface \texttt{GUIFactory} produces two products, namely \texttt{Button} and \texttt{Checkbox}. It's implementation is shown below:

\begin{lstlisting}[style=JavaStyle]
public class MacOSFactory implements GUIFactory {
    @Override
    public Button createButton() {
        return new MacOSButton();
    }
    @Override
    public Checkbox createCheckbox() {
        return new MacOSCheckbox();
    }
}
\end{lstlisting}

In this example, the class-level context identifies \texttt{GUIFactory} as an abstract \textit{creator} within the Abstract Factory pattern, while \texttt{MacOSFactory} serves as a \textit{concrete factory} producing platform specific product variants. The method-level context explains how each method instantiates a particular product (\texttt{MacOSButton}, \texttt{MacOSCheckbox}). The structural context emphasises the interface (implementation) relationship that supports polymorphic behaviour and extensibility. Such contextual understanding is essential for tasks such as design pattern identification and comprehension.


Table \ref{tab:dps_examples} shows a same examples of design pattern contextual summaries taken from the DPS corpus.

\begin{table}
    \centering
    \caption{Examples of generated summaries by DPS capturing the context}
    \renewcommand{\arraystretch}{1.2}
    \setlength{\tabcolsep}{5pt}
    \begin{tabular}{|c|p{5.9cm}|p{4.45cm}|}
        \hline
        \textbf{ID} & \textbf{Sample Design Pattern Code} & \textbf{DPS Summary} \\
        \hline
        1 & 
        \begin{minipage}{5.9cm}
        \begin{lstlisting}[style=JavaStyle]
public class EmailNotificationListener implements EventListener {
    private String email;

    public EmailNotificationListener(String email) {
        this.email = email;
    }

    @Override
    public void update(String eventType, File file) {
        System.out.println("Email to " + email + ": Someone has performed " + eventType + " operation with the following file: " + file.getName());
    }
}
        \end{lstlisting}
        \end{minipage} 
        & 
        \begin{minipage}[t]{4.45cm}
        \vspace{-4cm}
EmailNotificationListener acts as a concrete observer inside the Observer Pattern, which updates String parameter of eventType, File parameter of file. It is a public class that implements EventListener. It sends email notifications. The only method of EmailNotificationListener is void update.
        \end{minipage} 
        \\
        \hline
        2 & 
        \begin{minipage}{5.9cm}
        \begin{lstlisting}[style=JavaStyle]
public interface ComputerPart {
    public void accept(ComputerPartVisitor computerPartVisitor);
}
        \end{lstlisting}
        \end{minipage} 
        & 
        \begin{minipage}[t]{4.45cm}
        \vspace{-0.85cm}
ComputerPart acts as a visitor for element ComputerPart, which is inherited by ComputerPartDisplayVisitor, which accepts ComputerPartVisitor parameter of computerPartVisitor. It is a public interface. The only method of ComputerPart is accept (void). No methods call accept and accept calls no methods. 
        \vspace{0.2cm}
        \end{minipage}
        \\
        \hline
        3 & 
        \begin{minipage}{5.9cm}
        \begin{lstlisting}[style=JavaStyle]
public interface GUIFactory {
    Button createButton();
    CheckBox createCheckBox();
}
        \end{lstlisting}
        \end{minipage} 
        & 
        \begin{minipage}[t]{4.45cm}
        \vspace{-0.7cm}
GUIFactory acts as an abstract factory for Button, CheckBox, which is inherited by WindowsOSFactory, MacOSFactory. It is a public interface. The 2 methods of GUIFactory are createButton (Button) and createCheckBox (CheckBox). No methods call createButton and createButton calls no methods. No methods call createCheckBox and createCheckBox calls no methods. 
        \vspace{0.2cm}
        \end{minipage}
        \\
        \hline
        4 & 
        \begin{minipage}{5.9cm}
        \begin{lstlisting}[style=JavaStyle]
public class HtmlDialog extends Dialog {
    @Override
    public Button createButton() {
        return new HtmlButton();
    }
}
        \end{lstlisting}
        \end{minipage} 
        & 
        \begin{minipage}[t]{4.45cm}
        \vspace{-1.4cm}
HtmlDialog is a part of factory method pattern.  It is a public class that extends Dialog. It overrides the createButton() method to return an instance of HtmlButton. The only method of HtmlDialog is createButton (Button). createButton calls no methods. 
        \vspace{0.2cm}
        \end{minipage} 
        \\
        \hline
        5 & 
        \begin{minipage}{5.9cm}
        \begin{lstlisting}[style=JavaStyle]
public class BusAdapter implements Car {

  private Bus bus;

  public BusAdapter() {
    this.bus = new Bus();
  }

  @Override
  public void drive() {
    bus.run();
  }
}
        \end{lstlisting}
        \end{minipage} 
        & 
        \begin{minipage}[t]{4.45cm}
        \vspace{-2.4cm}
BusAdapter acts as an Adapter, which overrides driveCar. It is a public class that implements Car. It adapts the Bus class to use a Car interface. BusAdapter implements the Car interface and internally holds a reference to a Bus object.
The only method of BusAdapter is drive (void). No methods call drive and drive calls the run method. 
        \end{minipage}
        \\
        \hline
    \end{tabular}
    \label{tab:dps_examples}
\end{table}

\section{Experimental Setup}\label{sec:exper} This section describes the experimental setup of our study, including the construction of the dataset, the code features used for summarisation, the evaluation metrics applied, and the research questions that guide our analysis.

\subsection{Data Collection}\label{subsec:data}

\textcolor{blue}{To construct our corpus, we manually selected ten open-source educational repositories and five open-source general-purpose repositories from GitHub. The educational repositories were identified using the keyword \textit{“design patterns”} and were specifically created to illustrate or teach software design patterns by the authors, rather than to serve as general-purpose software projects used by software developers. Many of these repositories originate from academic contexts (e.g., university course materials) or are associated with instructional resources such as books and websites—for example, the widely referenced \textit{RefactoringGuru} repository.}

\textcolor{blue}{While some educational repositories included examples for all 23 Gang of Four (GoF) design patterns, others featured a broader or narrower subset. We manually curated each repository to retain only those directories relevant to the nine design patterns targeted in our study. This involved selecting example code files from appropriate directories and excluding test cases, non-Java files, and unrelated content.}

\textcolor{blue}{In contrast, the five general-purpose repositories—Eclipse JDT, Apache Hadoop, Spring Framework, Apache Camel, and jOOQ—are production-level projects that are actively maintained and used in industrial software development. They were selected to complement the educational examples with realistic, large-scale systems. We retained only those files that contained instances of the nine design patterns targeted in our study and excluded unrelated or non-Java files. The resulting dataset, referred to as the \textbf{DPS-Corpus}, now comprises slightly more than \textit{42,000} Java files.}

\textcolor{blue}{Table~\ref{tab:coll} presents the project names, their GitHub URLs, and the number of Java files containing implementations of the nine design patterns under investigation. The table summarises the distribution of design pattern instances across the fifteen Java projects included in our corpus. Each project demonstrates various design patterns, including \textit{Abstract Factory}, \textit{Adapter}, \textit{Decorator}, \textit{Facade}, \textit{Factory Method}, \textit{Memento}, \textit{Observer}, \textit{Singleton}, and \textit{Visitor}. The total number of pattern-related files per project ranges from 46 to 6,079. Among the general-purpose repositories, \textit{Apache Camel} contains the highest count while among the educational repositories, \textit{shihyu} remains the largest and \textit{shusheng007} has the lowest count with 46 files. Patterns such as \textit{Abstract Factory} and \textit{Factory Method} are widely implemented across both educational and general-purpose repositories, whereas patterns like \textit{Decorator} and \textit{Visitor} appear less frequently.}

\begin{table}[h]
\caption{Projects in the DPS corpus and the number of instances of each design pattern}
\label{tab:coll}
\centering
\scalebox{0.9}{
\begin{tabular}{
    p{5.5cm}|
    @{\hskip 8pt}r
    @{\hskip 8pt}r
    @{\hskip 8pt}r
    @{\hskip 8pt}r
    @{\hskip 8pt}r
    @{\hskip 8pt}r
    @{\hskip 8pt}r
    @{\hskip 8pt}r
    @{\hskip 8pt}r
    @{\hskip 8pt}|r
    @{\hskip 8pt}|r
    @{\hskip 8pt}|r
    @{\hskip 8pt}|r
    @{\hskip 8pt}|r
    @{\hskip 8pt}|r
}
\toprule
\textbf{Project \textit{(GitHub path)}} &
\rotatebox{90}{\textbf{Abstract Factory}} &
\rotatebox{90}{\textbf{Adapter}} &
\rotatebox{90}{\textbf{Decorator}} &
\rotatebox{90}{\textbf{Facade}} &
\rotatebox{90}{\textbf{Factory Method}} &
\rotatebox{90}{\textbf{Memento}} &
\rotatebox{90}{\textbf{Observer}} &
\rotatebox{90}{\textbf{Singleton}} &
\rotatebox{90}{\textbf{Visitor}} &
\rotatebox{90}{\textbf{Total}} \\
\midrule
RefactoringGuru \newline \vspace{0.3em} \textit{RefactoringGuru/design-patterns-java} & 9  & 4  & 5   & 8  & 6   & 13 & 5   & 2  & 7   & 59  \\
Iluwatar \newline \vspace{0.3em} \textit{iluwatar/java-design-patterns} & 13 & 4  & 4   & 5  & 7   & 3  & 11  & 5  & 8   & 60  \\
shihyu \newline \vspace{0.3em} \textit{shihyu/DesignPatternExample} & 141& 85 & 114 & 83 & 123 & 57 & 119 & 41 & 180 & 943 \\
saeidzabardest \newline \vspace{0.3em} \textit{saeidzebardast/java-design-patterns} & 14 & 4  & 7   & 5  & 5   & 3  & 5   & 1  & 7   & 51  \\
JamesZBL \newline \vspace{0.3em} \textit{JamesZBL/java\_design\_patterns} & 13 & 4  & 3   & 4  & 7   & 3  & 5   & 4  & 8   & 51  \\
Luisburgos \newline \vspace{0.3em} \textit{luisburgos/design-patterns} & 33 & 11 & 10  & 13 & 19  & 5  & 7   & 2  & 11  & 111 \\
shusheng007 \newline \vspace{0.3em} \textit{shusheng007/design-patterns} & 6  & 5  & 6   & 5  & 3   & 4  & 6   & 4  & 7   & 46  \\
AbdurRKhalid \newline \vspace{0.3em} \textit{AbdurRKhalid/Design-Patterns} & 10 & 4  & 5   & 8  & 6   & 3  & 5   & 1  & 7   & 49  \\
Quanke \newline \vspace{0.3em} \textit{quanke/design-pattern-java-source-code} & 10 & 13 & 10  & 5  & 7   & 3  & 9   & 4  & 8   & 69  \\
premaseem \newline \vspace{0.3em} \textit{premaseem/DesignPatternsJava9} & 13 & 7  & 9   & 6  & 10  & 8  & 7   & 5  & 4   & 69  \\
Spring Framework \newline \vspace{0.3em} \textit{spring-projects/spring-framework} & 114 & 81 & 23 & 510 & 146 & 20 & 26 & 76 & 19 & 1015 \\
Apache Camel \newline \vspace{0.3em} \textit{apache/camel} & 1342 & 51 & 319 & 1830 & 1399 & 100 & 159 & 109 & 2 & 6079 \\
Eclipse JDT \newline \vspace{0.3em} \textit{eclipse-jdt/eclipse.jdt.core} & 23 & 10 & 12 & 161 & 35 & 40 & 3 & 57 & 96 & 437 \\
Apache Hadoop \newline \vspace{0.3em} \textit{apache/hadoop/} & 72 & 9 & 20 & 1294 & 100 & 282 & 67 & 451 & 38 & 2333 \\
jOOQ \newline \vspace{0.3em} \textit{apache/hadoop} & 5	& 6 & 0 & 174 & 5 & 7 & 33 & 45 & 2 & 277 \\
\bottomrule
\end{tabular}
}
\end{table}

\subsection{Code Features}\label{subsec:features}
Inspired by the work on Design Pattern Detection by Nazar et al.~\cite{Nazar:2022}, where they employ code features to identify design patterns, we have used code features to build the summaries. In this study, we have increased the number of features from 15 to 20 and renamed some of the features used by Nazar et al~\cite{Nazar:2022}. While some features (and their names) remain the same, others, such as NOML, are newly introduced. Additionally, features such as class name (CN) and method name (MN) have been renamed to enhance clarity and understanding. These features can be categorised into basic, object-oriented, and binary features, and further subdivided into class-level and method-level features. Feature extraction is a crucial aspect of code analysis, and accurately identifying a design pattern necessitates thorough code analysis. Table~\ref{tab:features} lists the selected feature types, feature abbreviations, names, and descriptions.

The basic features capture the structural and descriptive elements of Java classes, methods, and fields. These include the names, modifiers, data types, and relationships within the source code. The binary features, on the other hand, represent boolean properties of the Java file regardless of its interface or other types, such as isInterfaceOrNot (ION) and isAbstractOrNot (AON), which classify the type of Java file. Object-oriented features such as ExtendsFrom (EXF) and ImplementsFrom (IMF) uncover inheritance and interface implementation details. Object-oriented features further capture inheritance and implementation relationships between classes and interfaces. Such relationships are needed to capture design patterns, as most of the GoF patterns use some sort of inheritance relationship. Together, these features enable a detailed analysis of code structures, dependencies, and design principles, which are the key essence of software design patterns.

These features are particularly valuable for studying and detecting design patterns, as they reflect fundamental object-oriented principles, i.e., encapsulation, inheritance, abstraction, and polymorphism. Analysing features such as method overrides (MethodOverride (MO)), incoming and outgoing method calls (IM, OM), and adherence to class hierarchies (EXF, IMF), helps in identifying the presence of design patterns. Furthermore, metrics such as MethodBodyLineType (MBLT) and NumberOfMethodVariables (NOMV) enable the identification of code smells and refactoring opportunities -- useful in identifying any variants of design patterns such as the Observer pattern which is often used in multiple ways such as Subscriber publisher model, as mentioned by Shvets \cite{Shvets:2021}, or the commonly used subject observer model. Using NOMV and MBLT, we can determine multiple variants of the same design pattern.

\begin{table*}[h]
\centering
\caption{The list of features with their types, names (with abbreviations), and descriptions}
\label{tab:features}
\begin{tabular}{p{0.5cm}p{4cm}p{6cm}}
\toprule
\textbf{Type} & \textbf{Feature Name} & \textbf{Description} \\ \midrule

\multirow{21}{*}{\rotatebox{90}{Basic}} 
 & ClassName (CN)                          & Name of the class such as WeatherObserver \\
 & ClassModifierType (CMT)                & Access Modifiers for classes (such as public, private etc) \\
 & ConstructorModifier (CM)               & Access Modifier of the constructor (public, private, static, etc) \\
 & FieldDataType (FDT)                    & Data type of the field such as int, String etc. \\
 & FieldModifierType (FMT)                & Access Modifier of the field (public, private, static, etc) \\
 & MethodName (MN)                        & Name of the method such as updateWeather. \\
 & MethodModifierType (MMT)               & Access Modifier of the method (public, private, static, etc) \\
 & MethodReturnType (MRT)                 & The Return Type of a method (void, int, Object etc) \\
 & MethodBodyLineType (MBLT)             & Type of code in a method’s body e.g. assignment statement, condition statements etc. \\
 & NumberOfMethodVariables or attributes (NOMV/NOMA) & Number of variables/attributes in a method. \\
 & NumberOfMethodCalls (NOMC)            & Number of method calls in a class. \\
 & NumberOfMethodLines (NOML)            & Number of lines in a method. \\
 & NumberofIncomingMethods (NIM)         & Number of calls to this method \\
 & NameofIncomingMethods (NaIM)          & Names of methods that call this method \\
 & NumberofOutgoingMethods (NOM)         & Number of methods a method calls \\
 & NameofOutgoingMethods (NaOM)          & Names of methods a method calls \\
 & ConstructorParameters (CP)            & List of Arguments of the Constructor \\
 & MethodParameters (MP)                 & List of Arguments of the Method \\
 & MethodOverride (MO)                   & Is Method using @Override or not \\
 & IncomingMethods (IM)                  & List of tuples containing Names of Methods called, and their classes \\
 & OutgoingMethods (OM)                  & List of tuples containing Names of Methods that call this class, and their classes \\
\midrule
\multirow{2}{*}{\rotatebox{90}{Binary}} 
 & isInterfaceOrNot (ION)                & A binary feature (0/1) if the Java file itself is an interface \\
 & isAbstractOrNot (AON)                 & A binary feature (0/1) if the Java file itself is an abstract class \\
\midrule
\multirow{2}{*}{\rotatebox{90}{OO}} 
 & EXtendsFrom (EXF)                     & List of classes that the Java file extends from, i.e. inherits \\
 & IMplementsFrom (IMF)                  & List of interfaces that the Java file implements \\ 

\bottomrule
\end{tabular}
\end{table*}

\subsection{Evaluation Metrics}\label{subsec:metrics}
We have used four evaluation metrics to verify the efficacy of our approach, namely BLEU-4, NIST, ROUGE-L, and Frugal Score. We describe them one by one in the subsections below.

\subsubsection{BLEU-4}\label{subsubsec:bleu}

The Bilingual Evaluation Understudy (BLEU) metric is widely used in natural language processing. It evaluates the quality of the outcome of generative tasks such as code comment generation \cite{Jiang:2017}, pull request descriptions \cite{Liu:2019}, and more. BLEU employs N-grams for matching and calculates the ratio of N-gram similarity between generated text and reference texts. In the context of summarisation, it compares an automatically generated summary or translation to one or more reference summaries or translations and calculates the n-gram overlap between them.

The BLEU score is calculated as the geometric mean of modified n-gram precisions, with a brevity penalty to account for overly short candidate summaries. The score is computed as follows:

\begin{equation}
\text{BLEU-4} = \text{BP} \times \exp \left( \sum_{n=1}^{N} \tau_n \log P_n \right)
\end{equation}

Where \( P_n \) is the ratio of the subsequences with length \( n \) in the candidate that are also in the reference. BP is the brevity penalty for short generated sequences, and \( \tau_n \) is the uniform weight \( 1/N \). We use corpus level BLEU-4, i.e., \( N = 4 \), as our evaluation metric since it has been demonstrated to be more correlated with human judgments than other evaluation metrics. The higher the BLEU score, the closer the candidate is to the reference. If the candidate is completely identical to the reference, the BLEU score becomes 100\%.

For our evaluation, we employed the \texttt{sentence\_bleu} function from the \texttt{nltk} package\footnote{\url{https://www.nltk.org/_modules/nltk/translate/bleu_score.html} accessed and verified on 31-03-2025} to compare the reference and generated summary sentences. The default weights of 0.25 for each n-gram (1-gram to 4-gram) were used to ensure a balanced contribution of different n-gram lengths to the final score. Additionally, we applied the built-in \texttt{method1} smoothing function to mitigate the issue of overly harsh scores, particularly in cases where non-matching n-grams would otherwise result in a score of 0. Before evaluation, both the reference and generated summary sentences were tokenised to facilitate accurate and consistent scoring.

\subsubsection{NIST}\label{subsubsec:nist}

NIST (National Institute of Standards and Technology) is a commonly used evaluation metric to assess the quality of automatically generated textual summaries. It compares an automatically generated summary to a reference summary and calculates various scores, such as unigram precision, unigram recall, and unigram F-measure \cite{Doddington:2002}. NIST is a variant of BLEU that places more emphasis on rewarding n-grams that are less frequent or more informative. It is calculated as follows:

\begin{equation}
\resizebox{\linewidth}{!}{$
\text{NIST} = \sum_{n=1}^{N} \left\{ \frac{\sum_{\text{all } n\text{-gram match that reference}} \text{Info}(n\text{-gram})}{\sum_{n\text{-gram reference}}} \right\} \cdot \exp(\beta) \log^2 \left[ \min \left( \frac{|\rho|}{|r|}, 1 \right) \right]
$}
\end{equation}

Where \( \beta \) is chosen to make the brevity penalty factor equal to 0.5 when the number of words in the system output is two thirds of the average number of words in the reference set. The \( \rho \) is the average number of words in a reference sentence. Chauhan et al.~\cite{Chauhan:2023} argue that NIST considers the information gained in each n-gram over the precision of n-grams, which implies that the NIST metric determines the importance of each n-gram in relation to the phrase. The importance of each n-gram is obtained with the help of weights. A correct n-gram will be given high weight if it is less likely to occur or is rare. The length penalty, known as the brevity penalty for the NIST score, is calculated such that small variations in the length of translation sentences do not significantly affect the overall score.

By assigning higher weights to n-grams, NIST encourages systems to generate summaries that contain novel and informative content. Compared to BLEU, the NIST metric does not penalize translation sentences for short length. Like BLEU, NIST is precision-oriented but avoids the pitfall of treating all n-grams equally, leading to a more significant and informative evaluation. The NIST metric (using the \texttt{nltk.translate.nist\_score} package\footnote{\url{https://www.nltk.org/api/nltk.translate.nist_score.html} accessed and verified on 31-03-2024}), an enhancement of the BLEU score, was used to assess the informativeness and precision of the summaries. We utilised the \texttt{sentence\_nist} function from the \texttt{nltk.nist} package module to compare the reference summaries with the generated summaries. To ensure meaningful and robust comparisons, the n-gram length was set to \( n = 4 \), thus avoiding excessively sparse matches and maintaining a balance between granularity and computational feasibility.

\subsubsection{ROUGE-L}\label{subsubsec:rouge}
The Recall-Oriented Understudy for Gisting Evaluation (ROUGE) metric is commonly used to evaluate the quality of machine-generated text. The ROUGE score compares an automatically generated summary with a reference summary and calculates the overlap between the two. ROUGE primarily measures recall, focusing on how much of the reference summary’s content is captured by the generated summary. 

ROUGE-L compares the longest common subsequence (LCS) between the generated summary and the reference summary. Unlike other ROUGE metrics that focus on n-gram overlaps, ROUGE-L captures the sequence of words in the summary that matches the reference, making it particularly useful for capturing the overall structure and fluency of the text. The ROUGE-L score is calculated using the following formula:

\begin{equation}
\text{ROUGE-L} = \frac{(1 + \beta) \times \text{Precision} \times \text{Recall}}{\text{Recall} + \beta^2 \times \text{Precision}}
\end{equation}

where Precision is the ratio of the length of the LCS to the length of the candidate summary. Recall is the ratio of the length of the LCS to the length of the reference summary. \( \beta \) is a parameter that balances the weight between precision and recall. Typically, \( \beta \) is set to 1, giving equal weight to precision and recall. Compared to BLEU and NIST, ROUGE-L is particularly strong at capturing the sequence and order of words, which makes it more aligned with human judgment of text coherence and fluency. Additionally, ROUGE-L balances both precision and recall, providing a comprehensive evaluation of how well the generated summary matches the reference in terms of content and structure. By focusing on the LCS, ROUGE-L is better at evaluating summaries that maintain the fluency and logical flow of the original text, which BLEU and NIST might miss.

To calculate the ROUGE-L score, we employed the Python-based ROUGE library\footnote{\url{https://pypi.org/project/rouge-score/}accessed and verified on 31-03-2025}. Within this library, we used the \texttt{rougeL} scorer, which implements the scoring mechanism based on the longest common sequence to evaluate the overlap between generated and reference texts. Additionally, we enabled the built-in stemmer functionality, which removes word suffixes to enhance the accuracy of word matching and improve the robustness of the evaluation. This approach ensures a reliable assessment of text similarity, particularly in terms of fluency and content overlap.

\subsubsection{FrugalScore}\label{subsubsec:frugal}

As the aforementioned metrics are designed for machine translation and measure n-gram similarity, i.e., a summary with similar meaning but differing sequences of n-grams would perform poorly, we employed another measure called FrugalScore as proposed by Phillips et al. \cite{Phillips:2022}. FrugalScore is a reference-based metric designed for evaluating Natural Language Generation (NLG) models. It provides a more efficient and cost-effective alternative to traditional metrics while retaining high performance. Compared to BLEU-4, ROUGE-L, and NIST, FrugalScore is more effective for several reasons:

\begin{enumerate}
    \item Traditional metrics like BLEU-4 and ROUGE-L have been criticised for their low correlation with human judgment. BLEU-4, for instance, was originally designed for machine translation and has been shown to have limitations when applied to other NLP tasks~\cite{Eddine:2022}. Similarly, ROUGE-L, while useful for summarisation, often fails to capture the significance of human language~\cite{Blagec:2022}. FrugalScore, on the other hand, has been designed to better align with human evaluators, providing a more accurate reflection of the quality of generated text~\cite{Eddine:2022,Blagec:2022}.
    \item BLEU-4 and ROUGE-L were developed with specific tasks in mind: translation and machine summarization, respectively. This makes them less versatile for other NLP tasks. FrugalScore offers a more flexible approach, making it applicable across a wider range of tasks without significant loss of accuracy \cite{Blagec:2022}.
    \item FrugalScore is designed to be computationally efficient, making it scalable for large datasets and real-time applications. This is a significant advantage over BLEU-4 and ROUGE-L, which can be computationally intensive and less practical for large-scale evaluations \cite{Eddine:2022}. Although our corpus is not large compared to other datasets, we still prefer to use FrugalScore to identify if it captures human sentiments in our machine-generated summaries better than other metrics.
    \item FrugalScore aims to provide a more holistic evaluation by considering multiple aspects of text quality, such as fluency, coherence, and relevance. This contrasts with BLEU-4 and ROUGE-L, which often focus on n-gram overlap and may miss other important quality dimensions \cite{Eddine:2022,Blagec:2022}.
\end{enumerate}

The exact formula for FrugalScore can vary based on its implementation and configuration, but a generalized representation is:

\begin{equation}
\resizebox{\linewidth}{!}{$
\text{FrugalScore} = \alpha \cdot \text{EmbeddingSimilarity} + \beta \cdot \text{LexicalSimilarity} + \gamma \cdot \text{TaskSpecificComponent}
$}
\end{equation}

Where EmbeddingSimilarity measures the semantic similarity between sentence embeddings of generated and reference texts. LexicalSimilarity quantifies surface level overlap (e.g., BLEU, ROUGE), and TaskSpecificComponent is an optional term to account for specific evaluation needs (e.g., fluency or diversity). \( \alpha \), \( \beta \), and \( \gamma \) are tunable weights for different components.

To compute FrugalScore, we used the official package available on GitHub\footnote{\url{https://github.com/moussaKam/FrugalScore} accessed and verified on 31-03-2025}. For the evaluation of our results, we employed the \texttt{mover-score} metric. This specific model leverages the BERT-based architecture, a transformer-based language model designed to capture the semantic meaning of text rather than relying solely on n-gram matching. The \texttt{mover-score} component of this metric quantifies semantic similarity using word embeddings to calculate the minimum distance required to align words between the hypothesis and the reference text. Given the computational demands associated with larger language models (LLMs), the medium-sized model was selected to ensure a balance between evaluation accuracy and computational efficiency.

\subsection{Research Questions}\label{subsec:RQ}
Our study investigates the following research questions:

\begin{enumerate}
    \item \textbf{RQ1: To what extent do automatically generated design pattern summaries align with human-written summaries when evaluated using standard metrics?}\\
    
    To address this question, we conducted a survey collecting human-written summaries for corpus files and performed a systematic comparison using established evaluation metrics, i.e., BLEU-4, NIST, ROUGE-L, and FrugalScore. The methodology and comparative analysis are detailed in Section \ref{subsec:metrics}.\\
    
    \item \textbf{RQ2: How do human evaluators perceive the quality of DPS summaries compared to human-authored summaries?}\\
    
    We designed and administered an expert evaluation survey via Prolific\footnote{\url{https://www.prolific.com/} accessed and verified on 08/03/2025}, an online research platform, to assess subjective quality ratings of both summary types. The experimental setup, participant criteria, and results are presented in Section \ref{subsec:quality}

\end{enumerate}

\section{Design Pattern Summariser (DPS) - Our Approach}\label{sec:dps}

The preliminaries and foundational concepts of our design pattern summariser (DPS) approach have been discussed in Section \ref{sec:exper}. In the following subsections, we detail the core components/modules of DPS, i.e., parsing the dataset, the design pattern checker, and summary generation modules.

\subsection{Parsing}\label{subsec:parse}
Upon collecting the data and identifying the code features, we construct an Abstract Syntax Tree (AST) based on the corpus. This AST encapsulates key structural elements including class names, method names, and other relevant features. For parsing the code, we employ the JavaParser API\footnote{\url{https://javaparser.org/} accessed and verified on 23/04/2025}, which processes the input corpus (DPS-Corpus) sequentially, file by file. JavaParser is a robust toolkit for analysing, transforming, and generating Java code. It delivers an AST representation of Java source code, enabling programmatic interaction and manipulation. Compatible with Java versions up to 17, JavaParser is widely adopted for tasks such as code analysis, refactoring, and automated code generation.

\begin{figure}[h]
\centering
  \includegraphics{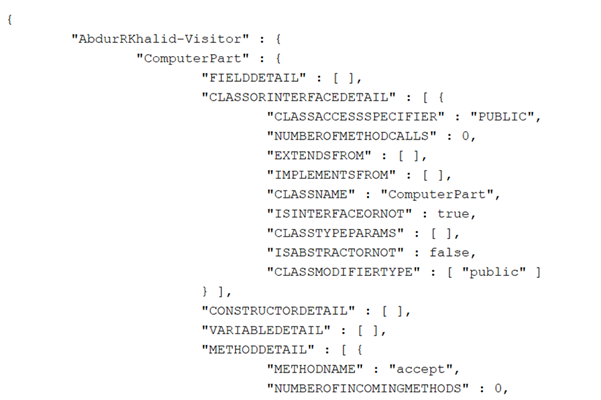}
\caption{JSON representation of the ComputerPart class from AbdurRKhalid project (alt text: JSON excerpt showing keys for classes, methods, and pattern roles used by DPS to generate summaries)}
\label{fig:json}       
\end{figure}

The extracted AST data (which contains the feature information as well) is stored in a structured dictionary and consolidated into a JSON file, which serves as the foundation for subsequent summary generation. This JSON representation -- depicted as a subset in Figure \ref{fig:json} -- constitutes the final output of this module. The figure illustrates an excerpt derived from the AbdurRKhalid\footnote{\url{https://github.com/AbdurRKhalid/Design-Patterns} accessed and verified on 08/03/2025} ComputerPart file, a component of the Visitor design pattern implementation.

\subsection{Pattern Checker}\label{subsec:pattern}

This module represents an improved iteration of our prior work, incorporating updated feature sets and heuristics-based algorithms to identify design patterns within the codebase. Unlike the existing version, which relied on a limited set of features, the current (our) approach leverages a broader range of features such as inheritance hierarchies, keyword usage, and other syntactic elements in conjunction with the constructed AST to verify whether the code matches the requisite pattern configuration.

Upon successful identification of a design pattern, the relevant information is embedded into the JSON representation (from the parsing phase). Our methodology aligns with the pattern structures defined by Shvets~\cite{Shvets:2021}, where we systematically search for specific feature combinations and their corresponding AST representations to validate the presence of the intended pattern.

For example, in the case of the Observer pattern, the algorithm verifies the existence of a subscriber publisher relationship. This entails ensuring that all concrete subscribers inherit from a subscriber interface and override the update method, as required by the pattern’s structural requirements. The process iterates until all essential code structures are confirmed, culminating in the formal identification of the design pattern.

\textcolor{blue}{It is important to note that DPS assumes the availability of a correctly detected design-pattern instance as input. The system operates downstream of established detection heuristics and focuses on summarising the structural and contextual features of these verified instances. While our pattern-checker module validates candidate instances using a 20-feature heuristic model, the goal of this component is not to propose a new detection technique but to ensure that summaries are generated from accurate pattern configurations. The summarisation quality therefore depends on the precision of the prior detection step.}

\subsection{Summary Generator}\label{subsec:summ}

We used SimpleNLG, a Java-based natural language generation API, to produce design pattern summaries. This framework leverages lexical resources and realisation components to construct grammatically correct English sentences from structured inputs. The JSON output generated by our Pattern Checker module serves as the input for this phase, with SimpleNLG transforming the structured pattern information into coherent natural language summaries.

\begin{figure}[h]
  \includegraphics[scale=0.44]{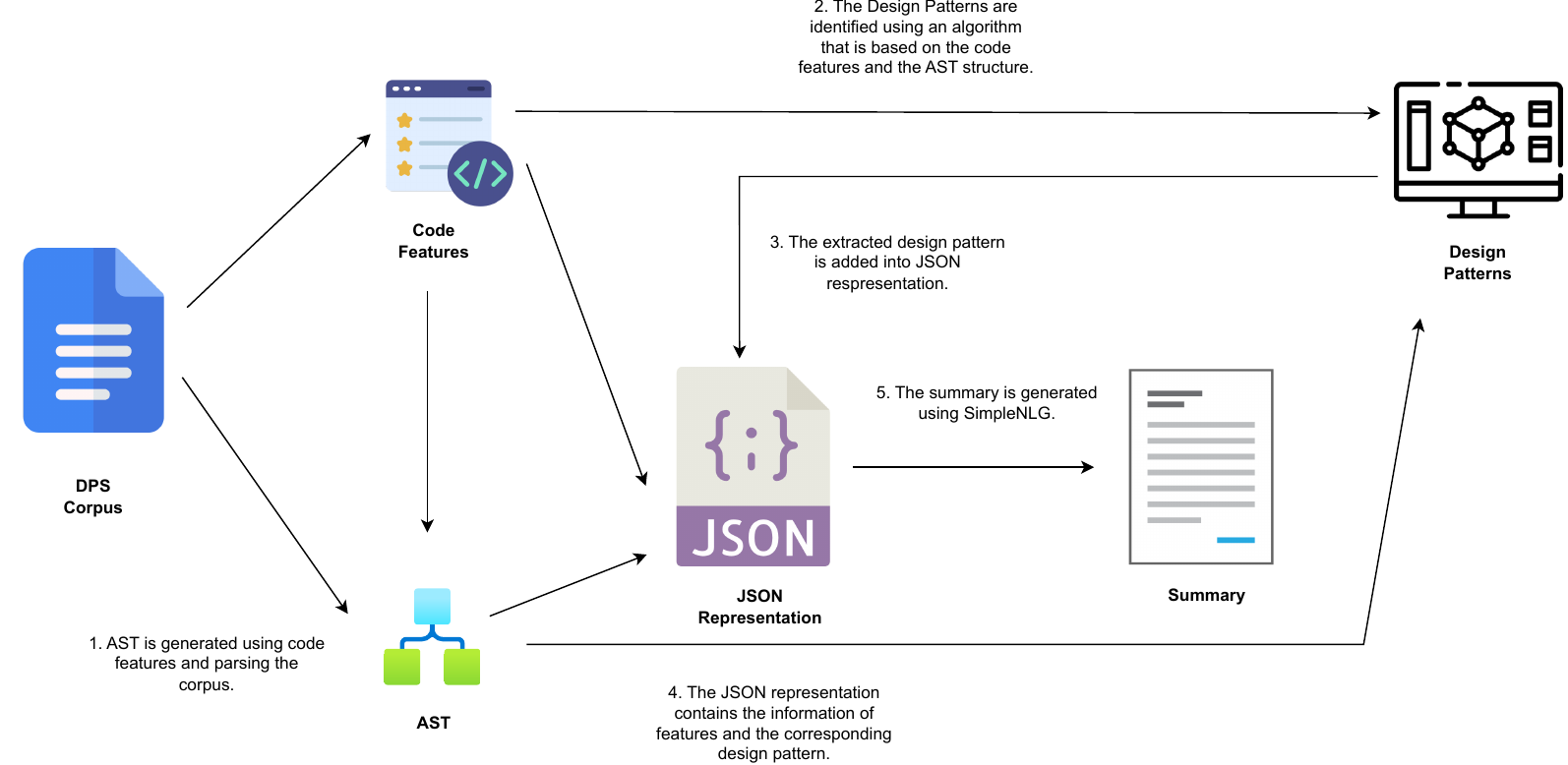}
\caption{DPS: Our design pattern summarisation approach (alt text: block diagram showing parsing to AST, pattern checker producing JSON, and summary generator producing natural-language descriptions)}
\label{fig:dps}       
\end{figure}

These automatically generated summaries effectively capture both the structural implementation and contextual usage of each design pattern within the dataset. The transformation process maintains the semantic relationships present in the original code while presenting them in linguistically appropriate form. Figure \ref{fig:dps} illustrates the complete architecture of our approach, demonstrating the integration between pattern identification and summary generation components.

\section{Empirical Evaluation}\label{sec:empir}
As discussed in Section~\ref{subsec:metrics}, to evaluate the efficacy of our approach, we compare our work with the summaries generated by humans.

\subsection{RQ1: Statistical Evaluation of DPS against Human Summaries}\label{subsec:stat}
We adopted a multi-stage approach to gather human written summaries for our study. Initially, we used the Prolific platform, implementing screening questions based on the methodology outlined in \cite{Danilova:2021} to collect qualified participants for writing summaries for the dataset. However, this approach proved ineffective, as most respondents either failed to meet our participation criteria or provided non-serious responses.

Consequently, we refined our recruitment strategy by targeting subject matter experts. We distributed our repository to postgraduate students and faculty members specialising in software engineering at Ada University (Azerbaijan). Additionally, we engaged faculty members from the University of Melbourne (Australia), securing meaningful participation from several qualified experts in the field. We set a criterion that the participants should have at least 3 years of experience using design patterns, have intermediate familiarity with the Java programming language, and belong to the software engineering major. Table~\ref{tab:dps_examples} shows five summaries of sample design patterns taken from the DPS-Corpus.

Our study engaged a cohort of 25 participants, consisting of postgraduate computer science students and software engineering faculty members as mentioned, to generate human-written summaries of the corpus. We required each code sample to be summarised by a minimum of three independent participants to ensure reliability. While doing so, we asked participants to identify the design pattern they think the code contains and then summarise it. During the annotation and summarisation of the code, certain disagreements arose among participants regarding the classification of pattern types. For instance, in some cases, participants exhibited a tendency to identify a pattern as an abstract factory, whereas the correct classification was a factory method. Thus, to verify the level of agreement between these participants we calculated Cohen's Kappa \cite{Carletta:1996} score, which was $>$ 0.7, i.e., there is moderate to high level of agreement between these participants.

Participants were given
flexibility in how they completed the task -- some summarised entire projects, while others focused on individual files. Each file was summarised by up to three participants. From these, we manually
selected one representative summary per file for use in our evaluation. The selected summaries were chosen based on clarity, accuracy, and completeness, with preference given to those that best
captured the design intent and usage context. On average, each participants summarised between 25-50 files, including both pattern implementations (code containing design patterns) and non-pattern files (such as client files which were part of the package). 
We compared DPS summaries with human summaries using four statistical measures discussed in Section \ref{subsec:metrics}. Table \ref{tab:eval} illustrates the values of BLEU-4, NIST, ROUGE-L and FrugalScore calculates as a result of comparison between both summaries. We calculate the score of each pattern in the directory and then take the average of it. The table shows the average score for each project


\begin{table}[h]
\centering
\caption{BLEU, NIST, ROUGE and FrugalScore scores for each of the projects when compared with DPS and human summaries}
\label{tab:eval}       
\begin{tabular}{lllll}
\hline\noalign{\smallskip}
Project & BLEU-4 & NIST & ROUGE-L & FrugalScore \\
\noalign{\smallskip}\hline\noalign{\smallskip}
RefactoringGuru & 0.51 & 6.6 & 0.4 & 0.56 \\
Iluwatar & 0.38 & 6 & 0.43 & 0.46\\
shihyu & 0.56 & 8.3 & 0.5 & 0.45 \\
saeidzabardest & 0.34 & 6.3 & 0.39 & 0.49 \\
JamesZBL & 0.46 & 6.6 & 0.41 & 0.48 \\
Luisburgos & 0.51 & 8.5 & 0.41 & 0.61 \\
shusheng007 & 0.39 & 5.8 & 0.42 & 0.45 \\
AbdurRKhalid & 0.38 & 6.8 & 0.44 & 0.55 \\
Quanke & 0.44 & 7.7 & 0.5 & 0.6 \\
premaseem & 0.42 & 7.2 & 0.48 & 0.51 \\
Spring Framework & 0.64 & 7.6 & 0.66 & 0.62 \\
Apache Camel & 0.62 & 6.9 & 0.67 & 0.63 \\
Apache Hadoop & 0.59 & 7.4 & 0.58 & 0.6 \\
Eclipse.jdt.core & 0.65 & 7.8 & 0.64 & 0.68 \\
jOOQ & 0.58 & 8.1 & 0.71 & 0.61\\
\textbf{\textit{Total (avg)}} & \textit{0.5} & \textit{7.17} & \textit{0.51} & \textit{0.55} \\
\noalign{\smallskip}\hline
\end{tabular}
\end{table}

The results demonstrate that DPS summaries achieve moderate to high quality across all measures -- notably, FrugalScore exhibited superior performance, showing the highest correlation with human judgments. This finding is particularly encouraging as it validates our system's ability to capture the contextual application of design patterns, a key objective of our research.

\subsection{RQ2: Quality of Summaries}\label{subsec:quality}
To further evaluate the quality of the DPS summaries, we asked human experts to rate the DPS generated summaries and human written summaries. We set the following screening criteria for the participants: they must be computer science graduates, must be fluent in English, must have a postgraduate computer science degree or higher and at least 3 years of experience in using software design patterns in industrial settings. Around 20 participants responded to the survey and 15 satisfied our criteria. The participants belonged to different demographics and fulfilled the screening criteria.

\textcolor{blue}{To mitigate potential bias, all survey participants were anonymised and were not informed which summary was human-written or automatically generated. Each code example was presented with two unlabeled summaries—Summary A and Summary B—and participants answered five quality questions without knowing the source of either summary. Participants were screened to ensure at least three years of professional experience using design patterns in industrial contexts, postgraduate-level computer-science education, and Java proficiency.}


We asked the following five survey questions:
\begin{enumerate}
    \item Which summary is more accurate?
    \item Which summary is more concise?
    \item Which summary content is more adequate?
    \item Which summary better captures the code context?
    \item Which summary better captures the design patterns?
\end{enumerate}
The participants were given three options to choose from for each of the questions: 1) machine generated summary, 2) human written summary and 3) both. 

Based on the results, we observe that human-written summaries were rated higher in terms of accuracy, conciseness, and adequacy---reflecting the thoughtful effort and domain knowledge of our expert annotators. At the same time, DPS-generated summaries outperformed in two dimensions: capturing the context of the code and effectively conveying the underlying design pattern. These are precisely the aspects our approach aims to automate, and the results are encouraging. Despite being fully automated, DPS is able to match or even surpass human summaries in the structural and contextual dimensions that matter most for design pattern understanding.

\textcolor{blue}{Table \ref{tab:qual} reports mean scores averaged across all participants for each evaluation dimension (accuracy, conciseness, adequacy, context captured, and pattern conveyed). These values represent the aggregated perception of quality rather than direct pairwise votes.}

\begin{table}[h]
\centering
\caption{Comparison of DPS and Human Summaries}
\label{tab:qual}       
\begin{tabular}{lll}
\hline\noalign{\smallskip}
 & DPS Summary & Human Summaries \\
\noalign{\smallskip}\hline\noalign{\smallskip}
Accurate & 4.5 & 4.5 \\
Concise & 4 & 5 \\
Adequate & 4 & 5 \\
Context captured & 5.5 & 4 \\
Convey DP & 5 & 3 \\
\noalign{\smallskip}\hline
\end{tabular}
\end{table}

Based on the results, we can conclude that the DPS summaries capture the design pattern context better than the human-generated summaries -- having higher average score.

\section{Discussion}\label{sec:discuss}

\textcolor{blue}{As shown in Table \ref{tab:qual}, the proposed Design Pattern Summaries (DPS) achieve accuracy on par with human summaries (4.5 each) while surpassing them in capturing contextual information (5.5 vs.4) and conveying the design pattern itself (5 vs.3), even though human summaries score higher in conciseness and adequacy.}

\textcolor{blue}{To illustrate these differences more concretely, we provide six representative examples of summary outputs: three instances where DPS-generated summaries outperform human-written ones, and three instances where human-generated summaries are judged to be more effective than those produced by DPS.}

\textcolor{blue}{Table~\ref{tab:good} shows three instances of DPS summaries performed better than human summaries.}

\begin{table}[ht]
\centering
\caption{\textcolor{blue}{Three instances of summaries where DPS performed better than Human Summaries.}}
\label{tab:good}
\begin{tabular}{|>{\centering\arraybackslash}m{0.5cm}|p{5cm}|p{5cm}|}
\toprule
\textbf{URL} & \textbf{DPS Summary} & \textbf{Human Summary} \\
\midrule
\rotatebox{90}{\url{EmailNotificationListener.java}} & 
EmailNotificationListener acts as a concrete observer inside the Observer Pattern, which updates String parameter of eventType, File parameter of file. It is a public class that implements EventListener. It sends email notifications. The only method of EmailNotificationListener is void update. & 
The EmailNotificationListener is a concrete observer as it receives notifications about events related to file operations. It overrides the update method that sends email notifications. \\
\midrule
\rotatebox{90}{\url{Dialog.java}} & 
The Dialog class acts as an abstract creator in the factory method pattern. Dialog acts as a factory method for Button, which is inherited by Dialog, It is a public abstract class. The 2 methods of Dialog are renderWindow (void) and createButton (Button). The createButton is overridden by all subclasses for button types for each type of dialog. No methods call renderWindow and renderWindow calls only one method : createButton method of class Dialog. Only one method (renderWindow method of class Dialog) calls createButton and createButton calls no internal methods. & 
The Dialog class is an Abstract Creator providing a template for creating dialogs that can generate buttons. \\
\midrule
\rotatebox{90}{\url{DefaultINodeAttributesProvider.java}} & 
DefaultINodeAttributesProvider acts as a caretaker for memento Attribute, which starts, which stops, which gets String[] parameter of pathElements, INodeAttributes parameter of inode Attributes. It has three methods start(), stop() and getAttributes(). It calls no method. & 
The DefaultINodeAttributesProvider class in Apache Hadoop HDFS provides a default implementation of INodeAttributeProvider. It includes empty start and stop methods and a getAttributes method that returns the provided INodeAttributes without modification, serving as a basic, no-operation attribute provider for HDFS inodes. \\
\bottomrule
\end{tabular}
\end{table}

	textcolor{blue}{Table \ref{tab:bad} shows examples where human summaries were better than DPS summaries.}

\begin{table}[ht]
\centering
\caption{\textcolor{blue}{Three instances of summaries where DPS underperformed than Human Summaries.}}
\label{tab:bad}
\begin{tabular}{|>{\centering\arraybackslash}m{0.5cm}|p{5cm}|p{5cm}|}
\hline
\rotatebox{90}{\textbf{URL}} & \textbf{DPS Summary} & \textbf{Human Summary} \\
\hline
\url{Computer.java} & 
Computer does not have any design pattern.  It is a public class that extends ComputerPart and that implements ComputerPart. The only method of Computer is accept (void). No methods call accept and accept calls no methods. & 
The Computer class represents a Concrete Element of Visitor, which allows operations to be performed on Mouse, Keyboard and monitor components by an external visitor. \\
\hline
\rotatebox{90}{\url{ScopedProxyBeanDefinitionDecorator.java}} & 
ScopedProxyBeanDefinitionDecorator does not follow any design pattern. It is a package-private class that implements BeanDefinitionDecorator. The only method is decorate(Node, BeanDefinitionHolder, ParserContext). No methods call decorate and decorate calls a utility to make a proxy. & 
The ScopedProxyBeanDefinitionDecorator class applies the Decorator design pattern within the Spring Framework’s AOP configuration context. It customises bean definition parsing by intercepting the <aop:scoped-proxy/> tag and wrapping the original bean with a scoped proxy. This enables context-aware proxying of Spring beans, particularly useful for injecting request or session-scoped beans into singleton scoped components while preserving AOP capabilities. \\
\hline
\rotatebox{90}{\url{Memeber.java}} & 
Member does not have any design pattern. It is a public interface. The only method of Member is getDescription (String). No methods call getDescription and getDescription calls no methods. & 
The Member interface defines a contract for ship members or related entities, requiring them to provide a description via the getDescription() method, supporting the Abstract Factory pattern's consistency. \\
\hline
\end{tabular}
\end{table}

\FloatBarrier

\textcolor{blue}{The quantitative results in Table \ref{tab:eval} reinforce that across 15 diverse projects, DPS achieved average scores of 0.50 for BLEU‑4, 7.17 for NIST, 0.51 for ROUGE‑L, and 0.55 for FrugalScore, indicating strong alignment with reference summaries. High NIST values for projects such as Luisburgos (8.5) and shihyu (8.3) highlight the method’s ability to capture informative n‑grams, while elevated ROUGE‑L scores for jOOQ (0.71) and Apache Camel (0.67) reflect substantial overlap in longer, semantically meaningful sequences. Altogether these results demonstrate that the proposed approach not only produces summaries competitive with human‑authored ones but also excels in preserving the contextual integrity of design patterns.}


\textcolor{blue}{To complement the subjective survey, we performed a task-based comprehension study to evaluate whether automatically generated summaries assist in understanding design-pattern roles within source code. A research engineer with three years of professional experience was asked to identify all participating roles for nine design patterns, both with and without summaries. Each pattern instance was selected randomly to avoid order effects. Across all patterns, the participant completed the tasks more quickly when summaries were provided (see Table \ref{tab:task}), while accuracy remained constant. This finding demonstrates that DPS summaries improve comprehension efficiency for realistic program-understanding tasks.}

\textcolor{blue}{For each design pattern, we randomly selected ten instances from the dataset. In each instance, the participant was provided with a GitHub link to a single class known to play one role in that pattern (e.g., the AbstractFactory class). The participant’s task was to identify all other classes corresponding to the remaining roles of that pattern. For example, for the Abstract Factory pattern, the participant had to find the AbstractFactory, ConcreteFactory, AbstractProduct, and ConcreteProduct roles.}

\textcolor{blue}{The study followed a within-subject design: for half of the instances (five per pattern), the participant was shown our automatically generated summary; for the other half, no summary was provided. All assignments were randomly selected to avoid order or learning effects. The participant was allowed to consult the Refactoring Guru definition (\url{https://refactoring.guru/design-patterns}) of each pattern during the task. Time to completion was self-recorded for each instance. Each instance was attempted once, and no outliers were excluded.}

\textcolor{blue}{We excluded any cases where fewer than three roles were present. Table 1 shows the average time (in minutes and seconds) required to complete the role-identification task, comparing the conditions with and without summaries.}

\begin{table}
    \centering
    \caption{\textcolor{blue}{Role Identification Summary results with and without summaries}}
    \label{tab:task}
    \begin{tabular}{ccc}
    \hline\noalign{\smallskip}
        Design Pattern & Without Summary  & With Summary\\\hline
        Abstract Factory & 2:26 & 1:39\\
        Adapter & 3:26 & 2:59\\
        Factory Method &  3:01 & 0.59\\
        Visitor & 2:23 & 1:48\\
        \noalign{\smallskip}\hline
    \end{tabular}
    
    \label{tab:placeholder}
\end{table}

\textcolor{blue}{Across all four patterns, tasks performed with summaries were completed more quickly than those without. Accuracy was consistent across conditions, indicating that the improvement was due to efficiency rather than guesswork or omission. These results suggest that the summaries helped the participant locate and recognize pattern roles more efficiently during program comprehension.}

\textcolor{blue}{Large language models (LLM) based tools such as ChatGPT have shown considerable potential in natural language generation; however, their performance in producing design pattern contextual summaries is inherently sensitive to prompt engineering. Prompt engineering introduces a substantial experimental dimension, requiring systematic design, tuning, and evaluation to ensure fairness and reproducibility. Incorporating such models would therefore expand the scope well beyond the objectives of this study. Our focus is instead on a domain specific, code feature driven approach that can be applied consistently without the variability introduced by prompt engineering dependent systems. This enables a more controlled evaluation of summary quality.}

\section{Threats to Validity}\label{sec:threats}
The section discusses the internal and external threats to the validity of the study.
\subsection{Threats to internal validity}\label{subsec:inter}
We have identified three threats to the internal validity of the study.
\begin{enumerate}
    \item The accuracy and reliability of this study are dependent on the correctness and robustness of the JavaParser. Any inherent defects or limitations in the parser may introduce inaccuracies in the representation or analysis of the source code. During the course of this research, a specific issue was identified wherein the parser occasionally failed to import files belonging to the same module, thereby disrupting interactions between files within the module. This issue, occasionally resulting in incomplete summaries for the affected modules, was resolved by implementing a custom SymbolSolver in JavaParser and reverting to a slightly outdated version of the parser to ensure stability and functionality.
    \item Furthermore, this study employs JSON representation for parsing and representing data. Errors in the JSON parsing process or inaccuracies in the data representation could compromise the integrity of the analysis and lead to misinterpretation of the results. As noted earlier, the aforementioned issue with the Java parser could also result in incomplete JSON representations. This challenge was mitigated by adopting the aforementioned adjustments to the JavaParser configuration.
    \item Lastly, the study uses Natural Language Generation (NLG) as a tool for generating natural language summaries. While functional, the generated summaries are relatively primitive in their current form. The quality of these summaries could potentially be enhanced by leveraging more advanced tools or techniques, such as SWUM (Software Word Usage Model) or grammar-based generators, which may offer more sophisticated and contextually accurate outputs. Future work could explore the integration of such tools to improve the overall quality and readability of the generated summaries.
    \item \textcolor{blue}{Our approach assumes that the design-pattern detections provided to the summariser are correct. Any inaccuracies or false positives in pattern detection may result in incomplete or misleading summaries. To mitigate this threat, we manually verified a sample of the detected pattern instances against reference implementations. Future work will explicitly quantify how detection errors affect summary quality and explore coupling the summarisation pipeline with probabilistic confidence estimates from the detector.}
\end{enumerate}
\subsection{Threats to external validity}\label{subsec:external}
The current study relies on a specific corpus of Java projects, which may limit the relevancy of the findings to other programming languages. A more generalised and diverse corpus, encompassing a wider range of programming languages, could enhance the applicability of the approach. Future work will aim to expand the corpus to include a broader variety of software projects to validate the approach across different languages.

In addition, the projects included in the corpus are all open-source and may not fully reflect the characteristics of proprietary or industrial codebases. Open-source projects often differ in coding style, documentation practices, and project structure compared to closed-source software. Furthermore, the corpus may underrepresent certain styles of Java development, such as Android applications or large-scale enterprise systems that use frameworks and architectures not present in the selected projects. These differences may limit the generalizability of the findings to broader software development contexts.

\section{Related Work}\label{sec:work}
In this section, we elaborate on two main threads of related work, including code summarisation and design pattern identification.
\subsection{Code Summarisation}\label{subsec:codes}

A significant amount of research has been conducted on code summarisation, with several new techniques and approaches being proposed to improve the accuracy and usefulness of the generated summaries. Code summarisation aims to generate brief natural language descriptions for source code. Automatic code summarisation approaches vary from manually crafted templates \cite{Sridhara:2010,Sridhara:2011,Mcburney:2014}, Information retrieval (IR) based \cite{Haiduc:2010,Haiduc2010a,Wong:2013} and learning based approaches \cite{Allamanis:2016,Iyer:2016,Movshovitz:2013}.

Creating manually crafted templates to generate code comments is one of the most common code summarisation approaches. Sridhara et al. \cite{Sridhara:2010} used a Software Word Usage Model (SWUM) to create a rule-based model that generates natural language descriptions for Java methods. Moreno et al. \cite{Moreno:2013} predefined heuristic rules to select information and generate comments for Java classes by combining them. These rule-based approaches have been expanded to cover special types of code artefacts such as test cases \cite{Zhang:2011} and code changes \cite{Buse:2010}. In general, these templates synthesise comments by extracting keywords from a given source code.

IR approaches are widely used in summary generation and usually search comments from similar code snippets. Haiduc et al.~\cite{Haiduc2010a} apply the Vector Space Model (VSM) and Latent Semantic Indexing (LSI) to generate term-based comments for classes and methods. Their work was replicated and expanded by Eddy et al. \cite{Eddy:2013} who exploited a hierarchical topic model. Wong et al. \cite{Wong:2015} applied code clone detection techniques to find similar code snippets using code comments. This work is similar to their previous work AutoComment \cite{Wong:2013}, which mines human-written descriptions for automatic comment generation from Stack Overflow.

In past years, some studies have attempted to provide natural language summaries using deep learning approaches. Iyer et al.~\cite{Iyer:2016} presented RNNs with attention to produce summaries that describe C\# code snippets and SQL queries. It takes the source code as plain text and models the conditional distribution of the summary. Allamanis et al. \cite{Allamanis:2016} applied a neural convolutional attentional model to the problem of summarising the source code snippets into short, name-like summaries. Hu et al. \cite{Hu:2018} leverage the API sequence within the source code to assist in the generation of code comments. Their approach TL-CodeSum fuses the source code and API sequences within it for better comment generation.
Nazar et al. \cite{Nazar:2016} identified summary sentences for code snippets from the web. They treated code snippets on the web as a natural language text and utilised textual features and machine learning ensembles to determine if the specific line in the code snippets could be a summary sentence. They incorporated crowdsourcing practices to gather textual features and annotate datasets. Their work was inspired by the earlier work of Ying et al. \cite{Ying:2013}, where they generated summaries of code fragments.
\subsection{Design Pattern Identification}\label{subsec:dp_id}
Identification of software design patterns is a constantly evolving research area, and new techniques are being developed to improve the accuracy and efficacy of identification methods. Nazar et al. \cite{Nazar:2022} has identified design patterns by employing source code analysis by building callgraphs and using source code features. They transformed the code into a verbose representation, applying the word2vec algorithm to that representation and using supervised classifiers to determine whether the Java files contain a specific pattern. Their classifier achieved more than 80\% Precision and outperformed state-of-the-art systems substantially in terms of Precision. They also used a web-based code annotation tool called CodeLabeller \cite{Nazar:2023} to label the corpus for the study, which is useful for supervised learning-based studies.

Relevant studies using code metrics in the identification of design patterns were conducted by Uchiyama et al. \cite{Uchiyama:2011}, who presented a software pattern detection approach using software metrics and machine learning techniques. They identified candidates for the roles that compose design patterns by considering machine learning techniques and utilising basic software metrics. In their book, Lanza and Marinescu \cite{Lanza:2007} discussed learning from the information extracted from design pattern instances that usually include variant implementations such as number of accessor methods. Fontana et al. \cite{Fontana:2011} introduced the microstructures that are considered the building blocks of design patterns. Thaller et al. \cite{Thaller:2019} has built a feature map for pattern instances using neural networks.

Many studies developed tools that use source code or its intermediate representation to identify patterns as well as machine learning models to predict patterns. Lucia et al. \cite{De:2011}, Tosi et al. \cite{Tosi:2009} and Zhang and Liu \cite{Zhang:2013} used static and dynamic analysis (or a combination of both) to develop Eclipse plugins that detect design patterns from source code. Moreno and Marcus \cite{Moreno:2012} developed a tool, \textit{JStereoType}, used for detecting low-level patterns (such as classes and interfaces) to determine the design intent of the source code. Zanoni et al. \cite{Zanoni:2015} exploited a combination of graph matching and machine learning techniques to implement a tool called \textit{MARPLE-DPD}. Niere et al. \cite{Niere:2002} designed the FUJABA Tool Suite which provides developers with support for detecting design patterns (including their variants) and design smells. Hautam’aki \cite{Hautamaki:2005}, in a thesis article, used a pattern-based solution and tool to teach software developers how to use development solutions in a project.

Some techniques (including tool generation) have applied reverse engineering to identify design patterns from source code and UML artefacts. For instance, Lucia et al. \cite{De:2011} built a tool using static analysis and applied reverse engineering through visual parsing of diagrams. Other studies such as Thongrak and Vatanawood \cite{Thongrak:2014}, Panich and Vatanawood \cite{Panich:2016} and Shi and Olsson \cite{Nija:2006} also applied reverse engineering techniques on UML class and sequence diagrams to extract design patterns using ontology. Brown \cite{Brown:1996} proposed a method for the automatic detection of design patterns by reverse engineering the SmallTalk code.
Several other approaches have exploited machine learning to solve the issue of variants. For example, Chihada et al. \cite{Chihada:2015} mapped the design pattern detection problem to a learning problem. Their proposed detector is based on learning from the information extracted from design pattern instances, which usually includes variant implementations. Ferenc et al. \cite{Ferenc:2005} applied machine learning algorithms to filter false positives from the results of a graph matching phase, thereby providing better precision in the overall output while considering variants. A recent work by Hussain et al. \cite{Hussain:2018} leveraged deep learning algorithms for the organisation and selection of DPs based on text categorisation. To reduce the size of training examples for DP detection, Dong et al. \cite{Dong:2008} proposed a clustering algorithm on decision tree learning to detect design patterns.

        extcolor{blue}{Recent advances in large language models (e.g., GPT-4, CodeLlama) have demonstrated strong general-purpose summarisation capabilities. Applying such models to design-pattern documentation is a promising direction for future work. However, because LLM outputs are sensitive to prompt phrasing, temperature, and model version, they are less reproducible and less interpretable than feature-driven approaches like DPS. Future studies could integrate both paradigms, combining explicit structural features with LLM-based contextual reasoning to balance precision and creativity.}

\section{Conclusion \& Future Works}\label{sec:conc}
Automatic summarisation has seen significant advancements in recent years, enabling the concise representation of texts, dialogues, audio, source code, and other similar software artefacts. Techniques for generation have ranged from basic template-based approaches to sophisticated machine learning methods. Despite these advancements, summarising software design patterns remains an underexplored area. In this initial study, we build on existing work on identifying design patterns using code features to create automated summaries of these patterns. The results show that our method effectively captures the context in which design patterns are applied within the code. Statistical analysis confirms that the generated summaries are closely aligned with human-authored ones, as evidenced by strong performance in metrics such as ROUGE, BLEU, NIST, and FrugalScore. A follow-up survey with expert participants shows that the DPS summaries capture the design pattern context better than the human written summaries.

In the future, we plan to experiment with our approach with large language model (LLM) methods using OpenAI or DeepSeek API to build natural language summaries of software design patterns or capture other aspects in code such as quality concerns. In addition, other summaries generation techniques such as SWUM or grammar-based generators could also be explored.

\begin{acknowledgements}
We would like to extend our gratitude to the individuals who have dedicated their time and effort to writing manual summaries of the code provided in the corpus and participated in the survey on the Prolific platform.
\end{acknowledgements}

\section*{Reproducibility}\label{sec:appendix}
For the purpose of reproducibility of our results, we have released our complete implementation with the annotated reference set and the result files to the public as an open-source project via our online appendix at \url{https://github.com/najamnazar/designpatternsummariser}. We have collected all projects in the corpus and have committed them to a GitHub repository which is accessible at \url{https://github.com/SamSike/DPS-Corpus}.

\appendix 

\section*{Appendix - Declaration}\label{sec:declaration}
\subsection*{Funding} This project was not supported by any internal or external funding.
\subsection*{Ethical Approval} Not Applicable
\subsection*{Informed Consent} All participants provided informed consent before taking part in this research.
\subsection*{Author Contributions} Najam Nazar was responsible for data collection, data analysis, experimentation (including code development), and manuscript writing. Sameer Sikka contributed to the experimentation and corrected bugs and errors. Christoph Treude was involved in manuscript writing and review.
\subsection*{Data Availability Statement} Not Applicable
\subsection*{Conflict of Interest} The authors have no conflicts of interest to declare relevant to this paper's content.
\subsection*{Clinical trial number} Not Applicable

%
%

\bibliographystyle{spmpsci}      
\bibliography{paper}


%
%

\end{document}